\documentclass{aa}
\usepackage{graphicx}
\begin{document}

\title{The basic parameters of $\gamma$-ray-loud blazars}
\author{J.H. Fan}

\offprints{{Dr. J.H. Fan}, \email{fjh@gzhu.edu.cn}}

\institute{1. Center for Astrophysics, Guangzhou University,
 Guangzhou 510400, China, e-mail: fjh@gzhu.edu.cn\\
 2. Chinese Academy of Science-Peking
University Joint Beijing Astrophysical Center(CAS-PKU.BAC),
Beijing, China \\ 3. National Astronomical Observatory, Chinese
Academy of Sciences, Beijing, China}

\date{Received Oct 25, 2003/ Accepted Jan. 11, 2005}

\abstract{ We determined the basic parameters, such as the central
black hole mass ($M$), the boosting factor (or Doppler factor)
($\delta$), the propagation angle ($\Phi$) and
 the distance along the axis to the site of
$\gamma$-ray production ($d$) for 23 $\gamma$-ray-loud blazars
 using their available variability timescales.
 In this
method, the absorption effect depends on the $\gamma$-ray
 energy, emission size and property  of the accretion disk. Using
the intrinsic $\gamma$-ray luminosity as a fraction $\lambda$ of
the Eddington luminosity, $L^{in}_{\gamma}=\lambda L_{Ledd.}$ and
the optical depth  equal to unity, we can determine the upper
limit of the central black hole masses. We found that the black
hole masses range between $10^{7}M_{\odot}$ and $10^{9}M_{\odot}$
when $\lambda$ = 0.1 and 1.0 are adopted. Since this method is
based on gamma-ray emissions and the short time-scale of the
sources, it can also be used for central black hole mass
determination of high redshift gamma-ray sources. In the case of
the upper limit of black hole mass  there is no clear difference
between BLs and FSRQs, which suggests that the central black hole
masses do not play an important role in the evolutionary sequence
of blazars. }

\maketitle

\titlerunning{The basic parameters of $\gamma$-ray loud blazars}
\authorrunning{J.H. Fan}

%------------------------------%
% Introduction Section
%%%%%-------------------------%%%

\section{Introduction}

The EGRET instrument at CGRO has detected many blazars (i.e. flat
spectrum radio quasars (FSRQs) and BL Lacertae objects (BLs)).
Blazars emit most of their bolometric luminosity in $\gamma$-rays
(E $>$ 100 MeV) (Hartman et al. 1999). Many $\gamma$-ray emitters
are also superluminal radio sources (von Montigny et al. 1995).
These objects share some common properties, such as luminous
$\gamma$-ray emission and strong variability in the $\gamma$-ray
and other bands on timescales from hours to days (see below).
These facts suggest that $\gamma$-ray emission in blazars is
likely arise from a jet. To explain its observational properties,
a beaming (black hole $+$ accretion disk $+$ jet) model has been
proposed. In the beaming model, a supermassive black hole is
surrounded by an accretion disk. Many authors have tried to
estimate the masses using different methods, (1) the reverberation
mapping technique (e.g. Wandel, Peterson \& Malkan 1999; Kaspi et
al. 2000), (2) the gas and stellar dynamics technique (see Genzel
et al. 1997; Magorrian et al. 1998; Kormendy \& Gebhardt 2001),
(3) the variability time-scale technique (Fan et al. 1999; Cheng
et al. 1999), (4) the broad-line width technique (Dibai 1984;
Wandel \& Yahil 1985; Padovani \& Rafanelli 1988; Laor 1998;
Mclure \& Dunlop 2001; Vestergaard 2000 based on the assumption
that the clouds in the broad-line region (BLR) are gravitationally
bound and orbiting with Keplerian velocities). The central black
hole mass is also found to be correlated with bulge luminosities
(Kormendy \& Richstone 1995), the bulge mass (Magorrian et al.
1998), the bulge velocity dispersion (Ferrarese \& Merritt 2000;
Gebhardt et al. 2000; Ferrarese et al. 2001) and the radio power
(Franceschini et al. 1998). However,  Ho (2002) found that the
radio continuum power either integrated for the whole galaxy or
isolated for the core is poorly correlated with the central black
hole mass. The tight $M_{\rm{BH}}-\sigma$ correlation can be used
for black hole mass determination. This relation is also used by
Barth et al. (2002) and Wu et al. (2002) for black hole mass
determination for Mrk 501 and other AGNs. Cao (2002) estimated the
black hole mass for a sample of BL Lacertae objects based on the
assumption that broad emission lines are emitted from clouds
ionized by the radiation of the accretion disk surrounding the
black hole.

Since there is a  large number of soft photons around the central
black hole, it is generally believed that the escape of high
energy $\gamma$-rays from the AGN depends on the $\gamma-\gamma$
pair production process. Therefore, the opacity of $\gamma-\gamma$
pair production in $\gamma$-ray-loud blazars can be used to
constrain the basic parameters. Becker \& Kafatos (1995) have
calculated the $\gamma$-ray optical depth in the X-ray field of an
accretion disk and found that the $\gamma$-rays should
preferentially escape along the symmetry  axis of the disk, due to
the strong angular dependence of the pair production cross
section. The phenomenon of $\gamma-\gamma$ "focusing" is related
to the more general issue of $\gamma-\gamma$ transparency, which
sets a minimum distance between the central black hole and the
site of $\gamma$-ray production (Bednarek 1993, Dermer \&
Schlickeiser 1994, Becker \& Kafatos 1995, Zhang \& Cheng 1997).
So, the $\gamma$-rays are focused in a solid angle, $\Omega =
2\pi(1-cos \Phi)$, suggesting that the apparent observed
luminosity should be expressed as $L_{\gamma}^{obs}= \Omega
D^{2}F^{obs}_{\gamma}(>100MeV)$, where $F_{\gamma}^{obs}$ and $D$
are observed $\gamma$-ray energy flux and luminosity distance
respectively. The observed $\gamma$-rays from an AGN require that
the jet almost points towards us and that the optical depth $\tau
$ is not greater than unity. The $\gamma$-rays are from a solid
angle, $\Omega$, instead of being isotropic. In this sense, the
non-isotropic radiation,  absorption and beaming (boosting)
effects should be considered when the properties of a
$\gamma$-ray-loud blazars are discussed. In addition, the
variability time scale may carry the information about the
$\gamma$-ray emission region. These considerations require  a new
method to estimate the central black hole mass and other basic
parameters of a $\gamma$-ray-loud blazar, which is the focus of
the present paper. In section 2, we introduce the method used to
estimate the black hole mass and three other parameters (the
Doppler factor, the propagation angle of the $\gamma$-rays and
their emission distance at the symmetric axis above the accretion
disk). In section 3, we present the discussion and a brief summary
of the paper.

H$_{0}$ = 75 km s$^{-1}$ Mpc$^{-1}$, and q$_{0}$ = 0.5 are adopted
throughout the paper.

\section{Mass estimation method and result}

\subsection{Method}

Here we describe our method of estimating the basic parameters,
namely, the central black hole mass ($M$), the boosting factor (or
Doppler factor) ($\delta$), the propagation angle ($\Phi$) and the
distance along the axis to the site of the $\gamma$-ray production
($d$) for $\gamma$-ray-loud blazars with short timescale
variabilities (see Cheng et al. 1999 for detail). To do so, we
consider a two-temperature disk (See Fig. 1). The $\gamma$-ray
observations suggest that the $\gamma$-rays are strongly boosted.
From the high energy $\gamma$-ray emission we know that the
optical depth of $\gamma$-$\gamma$ pair production should not be
larger than unity. In addition, the observed short-time scale
gives some information about the size of emitting region. This can
be used to constrain the basic parameters of a $\gamma$-ray-loud
blazar as in the following.

\begin{figure}
\vbox to7.2in{\rule{0pt}{7.2in}} \includegraphics{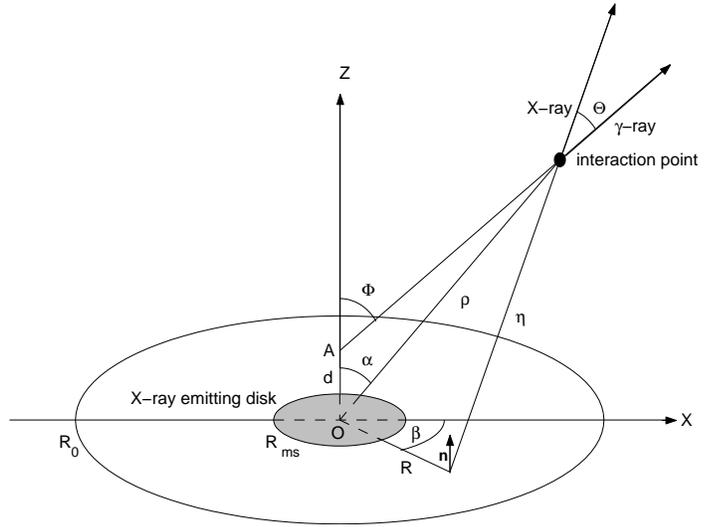} \caption{
Schematic diagram of $\gamma$-ray propagation above a
two-temperature disk surrounding a supermassive black hole.
$\gamma$-rays interact with the soft X-ray photons produced at all
points on the disk. The interaction angle between the $\gamma$-ray
and X-ray photons is $\Theta$, the angle between the $\gamma$-ray
trajectory and the z-axis is $\Phi$. $\eta$ is the distance
between the photon-photon interaction point and the soft photon
emission point in the disk. }
\end{figure}

{\it Optical depth\,\,\,}  Based on the paper by Becker \& Kafatos
(1995), we can obtain an approximate empirical formula for the
optical depth for a two-temperature disk case at an arbitrary
angle, $\Phi$ (see Cheng et al. 1999),

\begin{equation}
\tau_{\gamma\gamma}(M_{7},\Phi,d) = 9 \times
\Phi^{2.5}({\frac{d}{R_{g}}})^{-{\frac{2\alpha_{X}+3}{2}}}
+ kM_{7}^{-1}({\frac{d}{R_{g}}})^{-2\alpha_{X}-3}\;\;,
\label{tau}
\end{equation}
where $k$ is
\begin{eqnarray}
k=&4.61\times10^{9}{\frac{\Psi(\alpha_{X})(1+z)^{3+\alpha_{X}}F'_{0}(1+z-\sqrt{1+z})^{2}}{(2\alpha_{X}+1)(2\alpha_{X}+3)}}
\times\nonumber\\
&[{\frac{({\frac{R_{0}}{R_{g}}})^{2\alpha_{X}+1}-({\frac{R_{ms}}{R_{g}}})^{2\alpha_{X}+1}}{({\frac{R_{ms}}{R_{g}}})^{-1}-({\frac{R_{0}}{R_{g}}})^{-1}}}]({\frac{E_{\gamma}}{4m_{e}c^{2}}})^{\alpha_{X}}\;\;,
\end{eqnarray}
$M_{7}$ is the black hole mass in units of $10^{7}M_{\odot}$,
$\Psi(\alpha_{X})$ a function of the X-ray spectral index,
$\alpha_{X}$, $F'_{0}$ the X-ray flux parameter in units of
cm$^{-2}$~s$^{-1}$, $m_{e}$ the electron mass, $c$ the speed of
light, $R_{g}={\frac{GM}{c^{2}}}$ Schwarzschild radius,
$E_{\gamma}$ the average energy of the $\gamma$-rays. The inner
and outer radii of the hot region of a two-temperature accretion
disk (Becker \& Kafatos 1995) are $R_{0}$ and $R_{ms}$
respectively. Eq. (1) shows that the optical depth depends on $d$,
$\Phi$ and $M$.

{\it Time scale and the site of  $\gamma-$ray production\,\,\,}
The time scale gives us information  the emitting region (or the
distance along the axis to the site of  $\gamma$-ray production),
$d = {\frac{c \delta\Delta T_{D}}{1+z}}$. For convenience, we
express $d$ in the form of ($\Delta T_{D}$ in units of days).
\begin{equation}
{\frac{d}{R_{g}}} =1.73\times10^{3} {\frac{\Delta T_{D}}{1+z}}\delta
M_{7}^{-1}
\label{dis}
\end{equation}
where $\delta={\frac{1}{\Gamma(1-\beta cos\Phi)}}$ is the boosting
factor, ($\Gamma=(1-\beta^{2})^{-1/2}$ is the bulk Lorentz factor
and $\beta$ is the bulk velocity in unit of the speed of light
$c$.

{\it $\gamma-$Ray luminosity\,\,\,} In a relativistic beaming
model, the observed luminosity is correlated with the intrinsic
one in the frame comoving with the relativistic jet by
 $$L^{obs}_{\gamma}={\frac{\delta^{\alpha_{\gamma}+4}}{(1+z)^{\alpha_{\gamma}-1}}}L^{in}_{\gamma}$$
 where $\alpha_{\gamma}$ is $\gamma$-ray spectral index. As
mentioned above, the observed $\gamma$-ray flux,
$F_{\gamma}^{obs}(>100MeV)$, which is in units of
ergs~cm$^{-2}$~s$^{-1}$, can be expressed as a function of
 the intrinsic luminosity $L^{in}_{\gamma}$, the Doppler factor $\delta$,
 the luminosity distance $D$,
 and the solid angle $\Omega$ (or propagation angle $\Phi$)
$$F_{\gamma}^{obs}(>100{\hbox{MeV}})=(1+z)^{1-\alpha_{\gamma}}\delta^{\alpha_{\gamma}+4}L^{in}_{\gamma}/
\Omega D^2$$
  If we define  an isotropic luminosity as $L_{iso}=
4\pi~D^{2}F_{\gamma}^{obs} (>100MeV)$,
 we have
\begin{equation}
 L_{iso}^{45} =
{\frac{\lambda
2.52~\delta^{\alpha_{\gamma}+4}}{(1-cos\Phi)(1+z)^{\alpha_{\gamma}-1}}}M_{7}\;\;,
\label{lumi}
\end{equation}
where
 $L_{\gamma}^{in} = \lambda L_{Edd}=\lambda 1.26\times 10^{45}M_{7}$
is adopted, $\lambda$ is a parameter depending on specific
$\gamma$-ray emission models, $L_{iso}^{45}$ is the isotropic
luminosity in units of $10^{45}$ ergs~s$^{-1}$.

 From equation (\ref{lumi}), we can get the Doppler factor
  \begin{equation}
 \delta=(\frac{L_{iso}^{45}(1-cos\Phi)(1+z)^{\alpha_{\gamma}-1}}{\lambda 2.52M_{7}})^{\frac{1}{\alpha_{\gamma}+4}}
 \;\;.
 \label{delta}
\end{equation}

Substituting Eqs.  (\ref{delta}) into Eq. (\ref{dis}),
 we can
get a relation for $d(\Phi, M, L_{iso})$,
\begin{equation}
 d(\Phi, M, L_{iso}) = AR_{g}(1- cos \Phi)^{\frac{1}{\alpha_{\gamma}+4}} \;\;. \label{dis2}
\end{equation}
where
$$A = 1.73\times10^{3}\Delta
T_{D}(1+z)^{-\frac{5}{\alpha_{\gamma}+4}}M_{7}^{-{\frac{\alpha_{\gamma}+5}{\alpha_{\gamma}+4}}}({\frac{L_{iso}^{45}}{\lambda2.52}})^{{\frac{1}{\alpha_{\gamma}+4}}}$$

 Substituting Eqs.  (\ref{dis2}) and  (\ref{delta})
 into Eq. (\ref{tau}), we obtain a relation for
$\tau_{\gamma\gamma}(\Phi,M,L_{iso})$,
$$
\tau_{\gamma\gamma}(\Phi,M_{7},L_{iso}) =
 [9
 \times\Phi^{2.5}(1-cos\Phi)^{-\frac{2\alpha_{X}+3}{2\alpha_{\gamma}+8}}$$
\begin{equation}
  +kM_{7}^{-1}A^{-{\frac{2\alpha_{X}+3}{2}}}(1-cos\Phi)^{-\frac{2\alpha_{X}+3}{\alpha_{\gamma}+4}}]A^{-{\frac{2\alpha_{X}+3}{2}}}\;\;. \label{tau2}
\end{equation}
From the high energy $\gamma$-ray emission, we know that the
optical depth of $\gamma$-$\gamma$ pair production should not be
larger than unity. So, we can assume $\tau_{\gamma\gamma}
(\Phi,M_{7},L_{iso})=1.0$ for our purposes. However, there are two
variables in the equation of
$\tau_{\gamma\gamma}(\Phi,M_{7},L_{iso})=1.0$, so one should have
one more equation to determine the basic parameters. Fortunately,
the $\tau_{\gamma\gamma}(\Phi,M,L_{iso})$ shows a minimum for a
certain mass $M_{7}$ and angle $\Phi$. For a given mass, $M_{7}$,
the dependence  of $\tau_{\gamma\gamma}(\Phi,M,L_{iso})$ on $\Phi$
is illustrated in
 Figure 2, in which we show the case of 0208-512. For the source,
the relevant values are $\alpha_X$ = 1.04, $\alpha_{\gamma}$ =
0.69, $k=6.41$, $L_{iso}=2.0\times 10^{48} {\rm{ergs s^{-1}}}$,
$z=1.003$, $\lambda=0.1$, and $\Delta T$ = 134.4 hours
respectively.  In this sense, if we assume that the minimum value
of $\tau_{\gamma\gamma}(\Phi,M,L_{iso})$ is equal to 1.0, then we
will have the relation
${\frac{\partial\tau_{\gamma\gamma}}{\partial\Phi}}|_{M} = 0$.

Equation (\ref{tau2}) gives
\begin{eqnarray}
{\frac{\partial\tau_{\gamma\gamma}}{\partial\Phi}}|_{M} =
[22.5\Phi^{1.5}(1-cos\Phi) -
9\times{\frac{2\alpha_{X}+3}{2\alpha_{\gamma}+8}}\Phi^{2.5}sin\Phi&\nonumber
\\
-{\frac{2\alpha_{X}+3}{\alpha_{\gamma}+4}}kM_{7}^{-1}A^{-{\frac{2\alpha_{X}+3}{2}}}(1-cos\Phi)^{-{\frac{2\alpha_{X}+3}{2\alpha_{\gamma}+8}}}sin\Phi]\times
&\nonumber
\\
\times
[(1-cos\Phi)^{-{\frac{2\alpha_X+2\alpha_{\gamma}+11}{2\alpha_{\gamma}+8}}}A^{-\frac{2\alpha_{X}+3}{2}}],
\label{partau}
\end{eqnarray}
then, ${\frac{\partial\tau_{\gamma\gamma}}{\partial\Phi}}|_{M} =0$
suggests that
\begin{eqnarray}
22.5\Phi^{1.5}(1-cos\Phi) -
9\times{\frac{2\alpha_{X}+3}{2\alpha_{\gamma}+8}}\Phi^{2.5}sin\Phi&\nonumber
\\
-{\frac{2\alpha_{X}+3}{\alpha_{\gamma}+4}}kM_{7}^{-1}A^{-{\frac{2\alpha_{X}+3}{2}}}(1-cos\Phi)^{-{\frac{2\alpha_{X}+3}{2\alpha_{\gamma}+8}}}sin\Phi=
0 & \label{partau2}
\end{eqnarray}

\begin{figure}
\vbox to7.2in{\rule{0pt}{7.2in}} \includegraphics{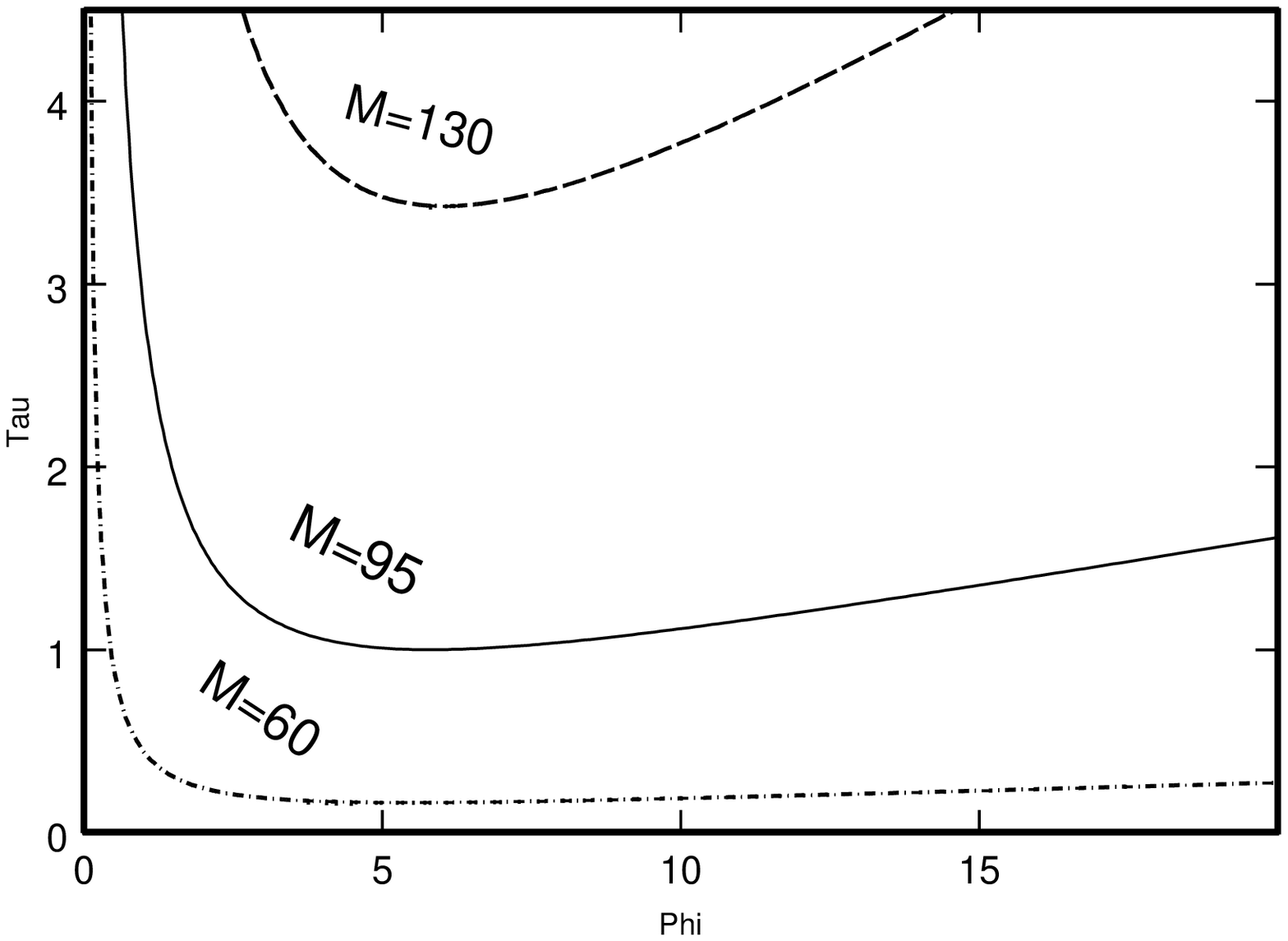} \caption{
Plot of the optical depth against the angle $\Phi$. For
illustration, we used the data of 0208-512,  the solid curve
stands for the $M_7$ = 95 case, while the dash-dotted curve for
$M_7$ = 60 and the dashed curve for $M_7$ = 130.  Here the
relevant values are $\alpha_X$ = 1.04, $\alpha_{\gamma}$ = 0.69,
$k=6.41$, $L_{iso}=2.0\times 10^{48} {\rm{ergs s^{-1}}}$,
$z=1.003$, $\lambda=0.1$, and $\Delta T$ = 134.4 hours
respectively.}
\end{figure}

Under this consideration, we can finally get four relations,
\begin{eqnarray}
{\frac{d}{R_{g}}} =1.73\times10^{3} {\frac{\Delta
T_{D}}{1+z}}\delta M_{7}^{-1} &\nonumber
\\
 L_{iso}^{45} =
{\frac{\lambda
2.52~\delta^{\alpha_{\gamma}+4}}{(1-cos\Phi)(1+z)^{\alpha_{\gamma}-1}}}M_{7}&\nonumber
\\
9 \times
\Phi^{2.5}({\frac{d}{R_{g}}})^{-{\frac{2\alpha_{X}+3}{2}}} +
kM_{7}^{-1}({\frac{d}{R_{g}}})^{-2\alpha_{X}-3} = 1 &\nonumber
\\
22.5\Phi^{1.5}(1-cos\Phi) -
9\times{\frac{2\alpha_{X}+3}{2\alpha_{\gamma}+8}}\Phi^{2.5}sin\Phi&\nonumber
\\
-{\frac{2\alpha_{X}+3}{\alpha_{\gamma}+4}}kM_{7}^{-1}A^{-{\frac{2\alpha_{X}+3}{2}}}(1-cos\Phi)^{-{\frac{2\alpha_{X}+3}{2\alpha_{\gamma}+8}}}sin\Phi=
0 &\label{relations}
\end{eqnarray}
in which, there are four basic parameters. So, for a source with
available data in the X-ray and $\gamma$-ray bands, the masses of
the central black holes, $M_{7}$, the Doppler factor, $\delta$,
  the distance along the axis to the site of the  $\gamma$-ray production, $d$, and the propagation angle with respect to the
axis of
 the accretion disk, $\Phi$, can be derived from Eq.
 (\ref{relations}), where $R_{ms} = 6R_{g}$, $R_{0}=30R_{g}$, and
$E_{\gamma}$= 1GeV are adopted.

\subsection{Results}

Since we are interested in the variability  timescale,  we present
here only the $\gamma$-ray-loud blazars with detected short time
scale variability. Since the variability timescale corresponds to
the different amplitude of variation for different sources and/or
different observational periods, we use the doubling timescale,
$\Delta T_{D} = (F_{minimum}/\Delta F) \Delta T$, as the
variability timescale, where $\Delta F = F_{maximum} -
F_{minimum}$ is the variation of the flux over the time $\Delta
T$. There are few simultaneous X-ray and $\gamma$-ray band
observations, so the data considered here are not simultaneous.
The $\gamma$-ray data by Hartman et al. (1999) are used to
calculate the $\gamma$-ray luminosity. The X-ray data are taken
from recent publications(See col. 5 in table 1). Except for the
two sources 1226+023 (Courviosier et al. 1988) and 2230+114 (Pica
et al. 1988), the doubling time is taken from Dondi \& Ghisellini
(1995).

The intrinsic $\gamma$-ray luminosity is unknown, so we assume
 it to be close to the Eddington luminosity, say $\lambda
L_{Edd.}$. In the present work, $\lambda$ = 0.1 and 1.0 are
adopted for the calculations. From the available X-ray and
$\gamma$-ray data, we can estimate the central black hole mass,
$M_{7}$ (see Table 1) and three other parameters ($\Phi$,
$\delta$, $d$, these values are not listed in Table 1). Since
short-term time scales and $\gamma$-ray emissions are included in
our consideration, only objects with those values can be involved
in the present paper. However, at present, short term timescales
are available only for 23 $\gamma$-ray loud blazars as listed in
Table 1, in which
 Col. 1 gives the name,
 Col. 2 the  redshift,
 Col. 3 the identification where Q stands for a flat spectral radio quasar
        and B for a BL Lacertae object,
 Col. 4 the 1 KeV X-ray flux density in units of $\mu$ Jy,
 Col. 5 the reference for Col. 4,
 Col. 6 the X-ray spectral index $\alpha_{X}$ (the averaged value of
        $<\alpha_{X}> = 0.67$ (Comastri et al. 1997) is adopted for
        FSRQs and $\alpha_{OX}=1.31$ is used for $\alpha_{X}$ for
        BLs if their X-ray spectral indices are unknown, as done by
        Ghisellini et al. (1998)),
 Col. 7 the reference for Col. 6,
 Col. 8 the flux F($>$100MeV) in units of $10^{-6}$ photon cm$^{-2}$ s$^{-1}$,
 Col. 9 the $\gamma$-ray spectral index $\alpha_{\gamma}$ = 1.0 is
        adopted for 0537-441 (see Fan et al. 1998); The data in col no.
        (7)
        and (8) are mainly from by Hartman et al. (1999)
        except for Mkn 501, which showed a flux of
        $F(>100\rm{MeV})=(0.32\pm0.13)\times10^{-6}$ photons cm$^{-2}$
        s$^{-1}$ with a photon index of 1.6$\pm$0.5 during a 1996
        multiwavelength campaign (see Kataoka et al. 1999),
 Col. 10 the doubling time scale in units of hours,
 Col. 11 reference for Col. 10,
 Col. 12 the observed isotropic luminosity in units of $10^{48}$erg~s$^{-1}$,
 Col. 13, the central black hole mass in units of $10^{7}M_{\odot}$ ($\lambda=1$),
 Co1. 14, the central black hole mass in units of $10^{7}M_{\odot}$
          ($\lambda=0.1$). The references in Table 1 (
  B97: Bloom et al. 1997;
 C97: Comastri et al. 1997;
 Ch99: Chiappetti et al. 1999;
 DG: Dondi \& Ghisellini 1995;
 F98: Fan et al. 1998;
 Fo98: Fossati et al. 1998;
 H96: Hartman 1996; H96b: Hartman et al. 1996;
 H: Hartman et al. 1999;
 K93: Kniffen et al. 1993; L98: Lawson \& McHardy 1998;
 M93: Mattox et al. 1993; M96: Madejski et al. 1996;
 M97: Mattox et al. 1997;
 P96: Perlman et al. 1996; P88: Pica et al. 1988;
 Q96: Quinn et al. 1996;
 S96: Stacy et al. 1996;
 U97: Urry et al. 1997;
 W98: Wehrle et al. 1998)

\clearpage
\begin{table*}
\caption{Black hole mass for 23 $\gamma$-ray-loud blazars}
\begin{tabular}{lccccccccccccc}
\hline\noalign{\smallskip}
 $Name$ & z& $ID$  & $f_{1KeV}$  &
 Ref &  $\alpha_{X}$ &  Ref &
 $ F$ &  $\alpha_{\gamma}$ &
  $\Delta T_{D}$ &  Ref &
 $L_{iso}^{48}$ &  $M_{1}$ &
 $M_{0.1}$ \\
 (1) &  (2)  &  (3)  &  (4)  &
 (5)  &  (6)  &  (7)  &  (8)  &
 (9)  &  (10)  &  (11) &  (12) &
 (13)  &  (14)\\
\noalign{\smallskip} \hline
 0208-512  & 1.003&Q& 0.61 & C97& 1.04
&C97 & 9.1   & 0.69 & 134.4 &S96 & 2.0  & 60.9 & 95.0\\

0219+428  & 0.444 &B& 1.56 & Fo98 & 1.6 & Fo98 & 0.25 &  1.01  &
30.0 & DG & 0.018 & 19.73 & 29.80\\

0235+164  & 0.94&B& 2.5&M96&1.01&M96&0.65&1.85& 72& M96& 2.0&
36.75& 53.69\\

0420-014 & 0.915 &Q& 1.08 & Fo98 & 0.67 & C97& 0.64 &  1.44  &
33.6 & DG & 0.175 & 10.98 & 16.43 \\

0458-020 & 2.286 &Q& 0.1 & DG & 0.67 & C97 & 0.68 &  1.45  & 144.0
& DG & 1.424 & 31.48 & 47.2\\

0521-365 & 0.055&B& 1.78 & DG & 0.68 & DG & 0.32&  1.63&  72& DG &
0.0002& 31.1 & 46.44\\

0528+134 & 2.07  &Q& 0.65 &C97& 0.54&C97 &3.08  & 1.21  & 24. &DG&
18.4 & 4.52 & 6.97 \\

0537-441 & 0.894 &B& 0.81 &C97& 1.16 &C97 & 2.0  & 1.0  & 16. &H96
& 3.01 & 10.46 & 15.96 \\

0716+714 & 0.3 & B&1.35& Fo98& 1.77 & Fo98& 0.46&  1.19&  1.92 &
DG & 0.013& 2.22 & 3.28 \\

0735+178 & 0.424 &B& 0.248 & Fo98 & 1.34 & Fo98& 0.30&  1.6  &
28.8 & DG& 0.013 & 20.03& 29.37\\

0829+046 & 0.18 &B& 1.07 & DG& 0.67 & C97& 0.34&  1.47 & 24. & DG&
0.003 & 12.14 & 18.2 \\

0836+710 & 2.17 &Q& 0.819 & Fo98 & 0.42 & Fo98& 0.33 &  1.62&  24.
& DG& 0.548 & 1.67 & 2.53 \\

1101+384 & 0.031&B& 37.33 & Fo98& 2.10& Fo98& 0.27& 0.57 &  1.92&
DG& 0.0001& 1.5& 2.31 \\

1226+023& 0.158 &Q& 12.07 & Fo98& 0.81& Fo98& 0.09&  1.58& 24 &
C88& 0.003& 8.9 & 13.27\\

1253-055 & 0.537 &Q& 2.43 &H96b& 0.68 &H96b & 2.8 & 1.02 & 12.
&K93 & 1.34 & 6.57 & 10.2\\

1253-055 & 0.538 &Q& 2.0 &L98 & 0.78 &L98 & 11.    & 0.97 & 6.
&W98 & 5.75 & 5.25 & 8.12\\

1510-089 & 0.361 &Q& 0.718 & Fo98& 0.90& Fo98& 0.49& 1.47& 57.6 &
DG & 0.018& 28.03& 41.9\\

1622-297 & 0.815 &Q& 0.08 &M97& 0.67 &C97 & 17.  & 0.87 & 4.85
&M97& 26.9 & 5.44 & 8.51 \\

1633+382 & 1.814 &Q& 0.42&C97 & 0.53&C97 &0.96  & 0.86 & 16. &M93
& 9.72 & 3.43 & 5.42 \\

1652+399 & 0.033 &B& 10.1  & C97 & 1.60&C97 &0.32 & 0.68 & 6. &Q96
& 0.0003 & 4.01 & 6.27 \\

2155-304 & 0.117&B& 0.058 & U97& 1.25&U97& 0.34& 0.56&  3.3 &
Ch99& 0.003& 3.26& 5.11\\

2200+420 & 0.07  &B& 1.84 &P96& 1.31 &P96 & 1.71 & 0.68 & 3.2 &B97
& 0.019 & 11.0 & 15.95 \\

2230+114 & 1.04& Q&0.486& Fo98& 0.67& C97& 0.51&  1.45&  48& P88&
0.184& 15.33 & 22.95\\

2251+158 & 0.859&Q& 1.08& Fo98& 0.62& Fo98& 1.16&  1.21&  1.92&
DG& 0.32& 0.57 & 0.87\\ \hline
\end{tabular}\\
 \end{table*}

\section{Discussion}

The central black hole plays an important role in the
observational properties of AGNs and has drawn much attention. It
may also shed some light on the evolution (Wang et al. 2001; Barth
et al. 2002; Cao  2002). There are several methods for black hole
mass determinations although consensus has not been reached. In
the present work, we proposed a new method to estimate the central
black hole mass. It is constrained by the optical depth of the
$\gamma-\gamma$ pair production and can be used if X-ray and
$\gamma$-ray emissions and short time-scale are known. This method
can be used to determine the central black hole mass of high
redshift gamma-ray sources.  It is an approximate empirical
method, which is obtained from the data/figures of Becker \&
Kafatos (1995).  The mass determined in the present paper
corresponds to an optical depth of unity and therefore the results
correspond to the upper limit of the central black hole mass. The
main difference between our consideration and others is that we
assumed that the $\gamma$-rays originate from a cone while others
think that the $\gamma$-rays are isotropic. However, our results
are consistent with those obtained by others as discussed  in the
following.

In our consideration, the estimated mass upper limits for a sample
presented here are in the range of $10^{7}M_{\odot}$ to
$10^{9}M_{\odot}$, $(0.57 \sim 60)\times 10^{7}M_{\odot}$ for
$\lambda = 1.0$ and $(0.87 \sim 95.0)\times 10^{7}M_{\odot}$ for
$\lambda = 0.1$. The real value of $\lambda$ will cause an
uncertainty in the mass, but the uncertainty caused by $\lambda$
will be negligible. When $\lambda$ decreases by a factor of 10,
the mass increases by a factor of $\sim$ 1.5. The results obtained
with the  present method are independent of the $\gamma$-ray
emission mechanism although it will depend the X-ray emission
mechanism.

For illustration, we will compare the mass estimation for two
sources, 3C 279 and Mkn 501. Several groups of authors have
estimated their masses.

{\it 3C 279\,\,\,} This quasar  displayed two outbursts, one in
1991 and another in 1996. The  upper limit of the central black
hole masses obtained from these outbursts are similar. For
$\lambda=1.0$ the masses were $M=6.57\times10^{7}M_{\odot}$ and
$M=5.25\times10^{7}M_{\odot}$ whereas for $\lambda=0.1$,
$M=10.2\times10^{7}M_{\odot}$ and $M=8.12\times10^{7}M_{\odot}$
for the 1991 and 1996 outbursts respectively. To fit the 3C 279
multiwavelength energy spectrum corresponding to the 1991
$\gamma$-ray flare, Hartman et al. (1996) used an accreting black
hole of $10^{8}M_{\odot}$; our result of $(6.57-10.2)\times
10^{7}M_{\odot}$ is consistent with their.

{\it Mkn 501\,\,\,} TeV and X-ray emission result shows that there
is a possible period od 23 days in the light curves of Mkn 501
(Hayashida et al. 1998; Kranich et al. 1999), which may suggest
binary black holes at  center (Rieger \& Mannheim 2000, 2003;
Villata \& Raiteri 1999). The black hole mass has  been determined
by many authors $(0.2-3.4)\times10^{9}M_{\odot}$ (Barth et al.
2002; Kormendy \& Gehbardt 2001; Merritt \& Ferrarese 2001; Rieger
\& Mannheim 2003). Our present result  shows that the central
black hole mass upper limit is $(4.01 \sim 6.27)\times
10^{7}M_{\odot}$. This result is lower than those claimed by the
authors mentioned above, but this difference is probably from the
facts that (1) the methods used to estimate the mass by different
authors are based on different assumptions, (2) there is a binary
black hole system at the center of the BL Lacertae objects, our
result corresponds to the less massive black hole. A binary black
hole system is successfully used in the explanation of the
periodic variability of the BL Lacertae object OJ 287 by Sillanpaa
et al. (1988), who claimed that the ratio of the secondary black
hole mass to the primary black hole mass is 0.004. Recently,
Komossa et al. (2003) reported that there is a binary black hole
system in NGC 6240. If there is a binary black hole system at the
center of Mkn 501, and the more massive black hole of $\sim
10^{9}M_{\odot}$ corresponds to the primary black hole and the
less massive black hole of $10^{7}M_{\odot}$ corresponds to the
secondary, then the ratio is 0.01, which is consistent with the
result for OJ 287 by Sillanpaa et al. (1988). In addition, our
result is consistent with that of  De Jager et al. (1999),
$(1-6)\times10^{7}M_{\odot}$.

BLs and FSRQs are two subclasses of blazars. From the
observational point of view, except for the emission line
properties (the emission line strength in FSRQs is strong while
that in BL is weak or invisible), other observational properties
are very  similar between them (Fan 2002). Their relationship has
drawn much attention (e.g. Sambruna et al. 1996; Scarpa \& Falomo,
1997; Ghisellini et al. 1998; D'Elia \& Cavaliere 2000; Bottcher
\& Dermer 2002; Fan 2002; Ciaramella et al. 2004). Bottcher \&
Dermer (2002) proposed that the accretion rate rather than the
central black hole masses play an important role in the
evolutionary sequence (from FSRQs evolving to BLs) of blazars. The
accretion rate in FSRQs is much larger than in BLs, there is not
much gas in  BLs to fuel the central black hole. In the present
paper, if we consider BLs and FSRQs separately, the distribution
of the mass upper limits is not very show much different (see Fig.
3), their average masses are log$M = 8.06\pm0.54$ for FSRQs, and
log$M = 8.13\pm0.46$ for BLs. There is no difference in black hole
mass between BLs and FSRQs. Recently, Wu et al. (2002) also found
that there is no mass difference between different subclasses of
AGNs. This result  suggests that the central black hole mass play
a less important role in the evolutionary sequence as pointed out
by Bottcher \& Dermer (2002). To investigate the evolutionary
process further, one should take into account the black-hole spin
and the accretion processes since the former effect is important
for non-thermal radio emission while the latter will change the
properties of the outflow.

\begin{figure}
\vbox to7.2in{\rule{0pt}{7.2in}} \includegraphics{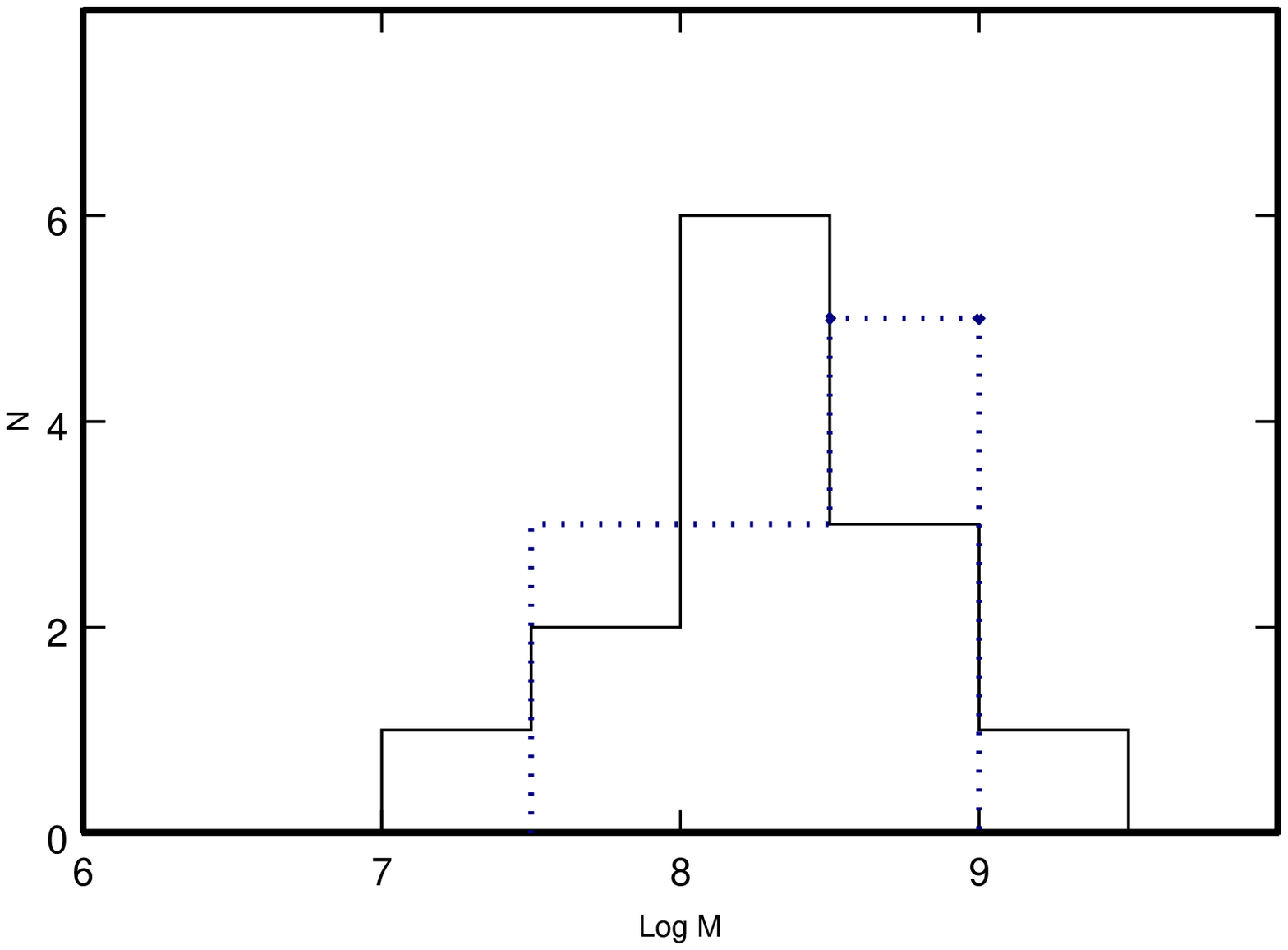} \caption{
Histogram of black hole mass for BLs(dotted lines)  and FSRQs
(solid lines) for the case of $\lambda=0.1$.}
\end{figure}

The high luminosity, rapid variability and superluminal motion
observed in some $\gamma$-ray-loud blazars suggest that the
$\gamma$-ray emission is  strongly beamed,  therefore, one can
assume that the beamed emissions arise from a certain solid angle.
In the present paper, our calculation show that the propagation
angle $\Phi$ and Doppler factor $\delta$ are in the range of
$9.^{\circ}68$ to 61.$^{\circ}14$ and 0.12 to 3.31 for $\lambda =
1.0$. For $\lambda = 0.1$ the value of $\Phi$ ranges from
$8.^{\circ}91$ to 56.$^{\circ}49$ and $\delta$ 0.16 to 4.6. In the
isotropic emission case, the $\gamma$-rays can be detected at any
angle, but in the case of non-isotropic emission, the emission is
 produced in the cone of a solid angle of $\Omega$, then
$\gamma$-rays will not be detected at any angle. From the
observational properties such as high luminosity, rapid
$\gamma$-ray variability  and superluminal motion, boosting
effects should be present in the $\gamma$-ray sources. So, the
optical depth should depend on the angle $\Phi$, the boosting
factor $\delta$, the central black hole masses  and the distance
along the axis to the site of  $\gamma$-ray production, $d$, which
can be expressed as $d(\Phi, M, L_{iso}) = AR_{g}(1- cos
\Phi)^{\frac{1} {\alpha_{\gamma}+4}}$. If the angle $\Phi$ is too
small, then $d$ is small, therefore, the site of the $\gamma$-ray
production is near  the center where the X-ray photon density will
be high. So, the optical depth is large. In this sense, the
smaller the angle $\Phi$, the smaller the $d$, and  the higher the
X-ray photon density, which results in a higher optical depth. On
the other hand, when the angle $\Phi$ is too large,  the boosting
factor $\delta$ is very small, so that the optical depth is also
large. Therefore, there are angles that correspond to a small
optical depth. From Fig. 2, one can see that the minimum value
depends on the central black hole mass, so one can choose a mass
so that the minimum value of the optical depth is 1.0. The mass
estimated in this way is an upper limit. From our calculation, we
found that Dopper factor $\delta <1$ in some cases; the lower than
unity Doppler factors do not conflict with the beaming argument
since we proposed that the emission is not isotropic
 in the present work. Other authors
assumed that the emission is isotropic.  The different assumptions
will result in a $(\frac{1-cos \Phi}{2})^{\frac{1}{4+\alpha}}$
times  difference in the Doppler factor. For  the $\gamma$-ray
emission regions, the obtained results indicate that they are in
the range of 17.2$R_{g}$ to 713$R_{g}$ ($\lambda = 1.0$) or
18.8$R_{g}$ to 640$R_{g}$ ($\lambda = 0.1$).

 In this paper, the optical depth of a $\gamma$-ray travelling in
 the field of a two-temperature disk and beaming effects have been used
 to determine the central mass, $M$,  for 23
 $\gamma$-ray-loud blazars with available short  time-scales.
 The masses obtained in the present paper are in the range of $10^{7}M_{\odot}$ to
$10^{9}M_{\odot}$ for the whole sample.  In the case of black hole
mass, there is no clear difference between BLs and FSRQs, which
suggests that the central black hole masses do not play an
important role in the evolutionary sequence of blazars.

%--------------------------------------------------------------------------
\begin{acknowledgements}

This work is partially supported by the  National 973 project
(NKBRSF G19990754), the National Science Fund for Distinguished
Young Scholars (10125313), and the Fund for Top Scholars of
Guangdong Province (Q 02114). I thank the anonymous referee for
the constructive suggestions and comments, Prof Jiansheng Chen and
Prof. Youyuan Zhou for suggestions, Dr. Alok C. Gupta for
reviewing the language for me, Dr. Hongguang Wang for useful
discussion. I also thank the Guangzhou City Education Bureau,
which supports our research in astrophysics and the Chinese
Academy of Sciences for the support for advanced visiting
scholars.

\end{acknowledgements}
%---------------------------------------------------------------------------

\end{document}